\documentstyle[aps,prl,multicol,epsfig]{revtex}

\begin{document}
\title{SINGLE PARTICLE SLOW DYNAMICS 
OF CONFINED WATER}
\author{Paola Gallo}
\address{Dipartimento di Fisica, Universit\`a di Roma Tre
and \\ Istituto Nazionale per la Fisica della Materia, 
Unit\`a di Ricerca Roma Tre,\\
Via della Vasca Navale 84
I-00146 Roma, Italy.}
\maketitle
\begin{abstract}
Molecular dynamics simulations of SPC/E water confined in a 
Silica pore are presented. The simulations
have been performed at different
hydration levels and temperatures to study the single-particle
dynamics. Due to the confinement and to the
presence of a hydrophilic surface,
the dynamic behaviour of the liquid appears to be 
strongly dependent on the hydration level.
On lowering temperature and/or 
hydration level the intermediate scattering function displays 
a double-step relaxation behaviour
whose long time tail is strongly non-exponential.
At higher hydrations two quite distinct subsets
of water molecules are detectable. Those belonging to the first two
layers close to the substrate suffer a severe slowing down already at 
ambient temperature. 
While the  behaviour of the remaining ones is more resemblant to
that of supercooled bulk SPC/E water. 
At lower hydrations and/or temperatures 
the onset of a  slow dynamics due to the cage effect
and a scenario typical of supercooled liquids approaching the 
kinetic glass transition is observed.
Moreover, for low hydrations and/or temperatures, the intermediate
scattering function clearly displays an overshoot,
which can be assigned to the so called ``Boson Peak''.
\end{abstract}

\begin{multicols}{2}

\section {INTRODUCTION}

Water plays a most fundamental role on earth and its anomalous
properties as a function of pressure and temperature are the subject of a
longstanding scientific debate. Nevertheless there are still 
many crucial questions  that remain to be answered.
Some of the most important are described throughout all 
the present issue that is dedicated to 
metastable water \cite{pablo}.
In particular, the supercooled region 
has attracted a wide attention in the last few years.
Thermodynamic as well as transport, dynamic, properties are not
completely understood in this region.
The main reason is that experimentally water can be supercooled only down to 
$T_H\sim236K$. 
Then, because of homogeneous nucleation 
due to impurities, the liquid is driven toward 
the crystal phase \cite{angell}. Therefore studies of the liquid 
both  through  classic computer Molecular Dynamics, 
MD, and through theoretical models
offer us ways to penetrate deep inside that region  
in the attempt to release the information that is 
not directly accessible to experiments.

In particular the normal limit of supercooling and the possibility 
of vitrification of water is a fundamental  point that necessitates to be 
clarified. A recent experiment sustains the hypothesis that 
the amorphous phase can be connected to the normal liquid phase
through a reversible thermodynamic path~\cite{speedy}.  
MD simulations of bulk supercooled SPC/E~\cite{BGS87} water showed
a kinetic glass transition  as predicted by 
mode coupling theory (MCT) \cite{goetze}
at a critical temperature $T_C\sim T_S$ \cite{gallo}, where $T_S$
is the singular temperature of water~\cite{SpeedyAngell}, 
which is $T_S=228K$ or, for SPC/E, 
$49$ degrees below the temperature of maximum density. 

Within this framework a
comparison of the behavior of the bulk liquid with the same liquid
in a confined environment is highly interesting 
since both the phase diagram and the dynamic
behaviour of confined water could present 
an analogy with the same liquid in the bulk phase.
The outcomings of this kind of researches can
have therefore important experimental implication for the bulk. 
In particular experimentally forbidden regions of the phase 
diagram of bulk water could become accessible through the study of 
confined water.
On the other hand a study of the modification of confined water with respect
to the bulk is highly interesting 
for the development of both biological and industrial applications.

It has been inferred from experiments, although not conclusively
proved, that confining water could be equivalent to the supercooling 
of the bulk \cite{chen,chenla}.
In this paper I will inquire on how far this analogy can be pushed
for SPC/E water confined in a silica pore and its bulk phase.

Among the different systems studied experimentally 
water confined in porous Vycor glass is one of the most interesting
with relevance to catalytic processes and enzymatic
activity. Vycor offers in fact to water a well characterized network of
cilidrical pores, is composed of simple molecules $SiO_2$, 
has a quite well characterized structure
with a sharp distribution of
pore sizes with an average diameter of $\sim 40 \pm 5 \AA $,
i.e. the same order of magnitude of many biological confining environments,
and a strongly hydrophilic inner surface.
For these reasons several experiments on water-in-Vycor
have been performed \cite{chenla,chen+mc,mar1}.

I shall show in the following the results concerning
the single particle dynamics obtained from a series of MD simulations 
of SPC/E water confined in a cylindrical silica cavity
\cite{spohr,gallo1,spohr2,gallo2}. 
The pore has  been modeled 
to represent the average properties of Vycor pores.
The simulated system from one side represents therefore 
a rather general confining environment 
and from the other 
offers the possibility of a direct comparison with experiments.
The results shown in the following are focused on the role of hydration level
on slow dynamics of confined water in comparison with the bulk 
properties upon supercooling.

In the next section the details of the model of the pore
are described together with the simulation details. 
In the third section the single particle 
dynamics of confined water is discussed in comparison with that of bulk 
SPC/E water. The last section concludes the paper with a discussion
on the results.

\section {MODEL AND SIMULATION DETAILS}

A single pore has been constructed for the simulations since in the
range of timescales involved in this kind of dynamics water does not leak
out of the pore.   
A cubic cell of vitreous $SiO_2$ was built by melting a $\beta$-cristobalite
single crystal and then by quenching the system down to ambient temperature 
\cite{brodka,feuston}. 
The side of the cube is $d\sim 70$~$\AA$. 
Inside the cell  a cylindrical cavity, the pore,
with diameter of $40$~$\AA$ was created by removing all the atoms
lying within a distance of $R=20$ $\AA$ from the z-axis passing through the
center of the cube. 
The inner surface of the pore was then ``corrugated'' 
by removing from it all the $Si$ atoms which were bonded to less than four 
$O$. At the surface oxygen atoms can be classified as bridging oxygens (bO)
if they are bonded to at least two silicon atoms, non-bridging oxygens
(nbO) otherwise. Then the dangling bonds of the nbO were saturated by 
attaching an hydrogen to each nbO. This procedure mimics the experimental
preparation of the sample of water-in-Vycor
in which internal surfaces of dessicated Vycor
are hydrogenated before  hydration.
At the end of this process the cell contains $6400$ silica atoms,
$12500$ bO and $230$ nbO. This last number yields a surface density of
acidic hydrogen of $2.5$ $nm^{-2}$ in good agreement with the $2.3$ $nm^{-2}$ 
obtained in the experiments \cite{hirama}.   
The simulations have been  performed in the NVE ensemble introducing 
water molecules into the pore.
Water molecules interact among themselves via a SPC/E potential, 
and with Vycor atoms via an effective pair potential 
\cite{brodka}.  
Both potentials have a Coulombic part plus a Lennard Jones term 
between oxygens. The parameters of the simulation are reported elsewhere
\cite{spohr,rovere}. 
The interactions cutoff is $9$ $\AA$ and 
the shifted force method has been used.
For the Coulomb interaction it was checked that 
with respect to the correct use of the Ewald summation this technique
was not producing qualitatively different results
\cite{spohr}.
The timestep of the simulations was 2.5 fs. Equilibration, reached through
a  Berendsen thermostat \cite{beren}, was monitored via the time dependence 
of the potential energy and the total energy.
Along the pore axis (z-direction) periodic boundary conditions are 
applied to the water molecules. 
Along the remaining directions (xy) water is confined. 
The glass is rigid. 

\begin{figure}[t]
\centering\epsfig{file=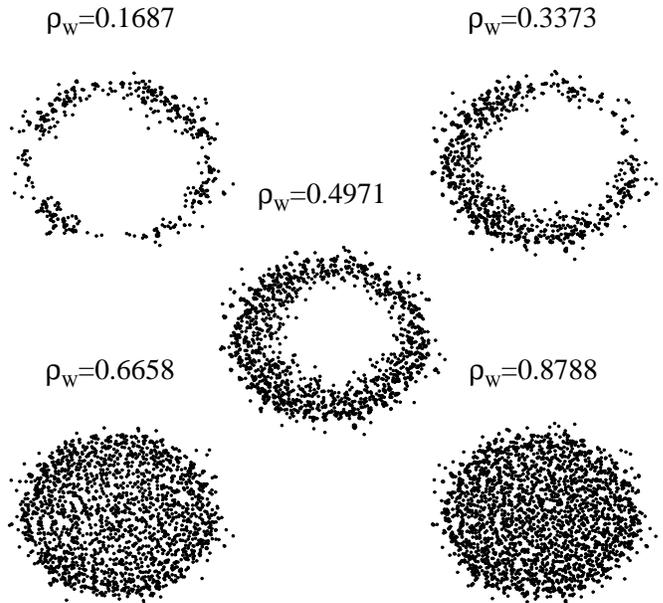,width=1\linewidth}
\caption{ 
Snapshots from simulations of SPC/E water inside a cylindrical
cavity of diameter 40 \AA{} at ambient temperature. 
Only oxygens are shown. $\rho_W$ is the global water density in $g/cm^3$
(see Table \protect\ref{tab1}).}
\protect\label{fig:1}
\end{figure} 

The molecular dynamics calculations have been performed for different
numbers of water molecules, corresponding to different levels of
hydration of the pore. The definition of hydration level of the pore deserves
some consideration. Experimentally Vycor glass absorbs water up to $25\%$
of its dry weight, this is defined as equilibrium or full hydration 
($h_f\simeq0.25$ g of water/ g of Vycor). 
The experimental density of water when confined in such a pore 
at full hydration is estimated to be $11\%$ less than its bulk value
at ambient conditions i.e. $\rho_W=0.0297$ molecules/$\AA^3=0.8877g/cm^3$ \cite{Ben}.
A partially hydrated sample is then experimentally 
obtained by absorption of water in the 
vapor phase until the desired level of hydration is reached.
For the designed pore of this simulation the 
density of the full hydration is obtained for 
$N_W = 2661$, where $N_W$ is the number of water molecules introduced in
the pore. 
The five hydration levels investigated in the present work are 
reported in Table \ref{tab1}.
I will discuss here the results obtained for two temperatures:
$T=298$~$K$ and $T=240$~$K$. 

\begin{table}[b]
  \caption{\label{tab1}
    Hydration levels of the pore. $N_W$ is the number of water
    molecules and $\rho_W$ the corresponding global density.
   $^{(a)}$: based on estimated value for full hydration
   N$_W =$ 2661  molecules (see text).}
    \begin{tabular}{ccccc}
    N$_W$ & \% hydration$^{(a)}$ & $\rho_W$ ($g/cm^3$) \\ \hline
    500  & 19\% & 0.1687  \\ 
    1000 & 38\% & 0.3373  \\
    1500 & 56\% & 0.4971  \\
    2000 & 75\% & 0.6658  \\
    2600 & 99\% & 0.8788  \\ 
  \end{tabular}
\end{table}

\begin{figure}[t]
\centering\epsfig{file=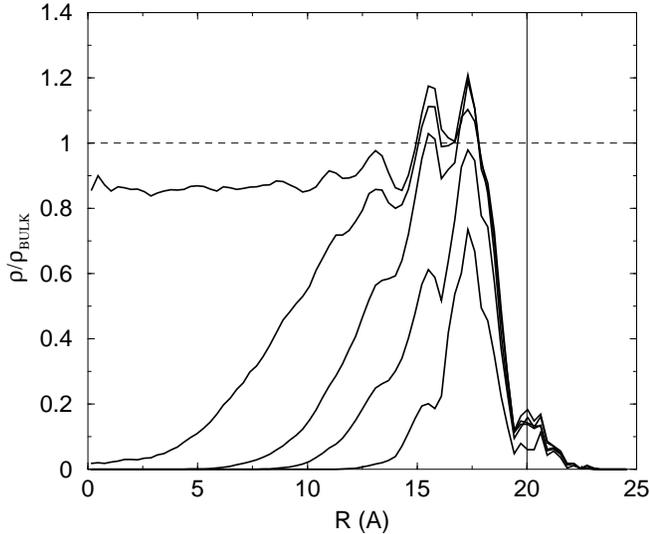,width=1\linewidth}
\caption{Radial density profiles 
normalized to the bulk 
for oxygen atoms at ambient temperature 
for the five hydration levels investigated:
first curve on the top corresponds to $\rho_W=0.8788g/cm^3$,
last on the bottom to
$\rho_W=0.1687g/cm^3$. The vertical line schematically indicates the pore
surface. Right inside the pore the density is
different from zero because few molecules are trapped in the pockets
of the corrugated pore surface (see text).}
\protect\label{fig:2}
\end{figure} 

The snapshots, see Fig.1, of the system  at ambient
temperature show that the simulated surface
is strongly hydrophilic. At all the degree of hydration water adsorbs
on the Vycor surface. A wide variety of water cluster
is visible for the lowest global density investigated 
$\rho_W=0.1687$ $g/cm^3$, which is lower than the estimated  
monolayer coverage  ($\rho_W\simeq 0.2219$ $g/cm^3$). 

The density profiles of the oxygen atoms of confined
water are shown in Fig.2.
It is observed already at lower hydrations the
presence of a layer of water molecules wetting the substrate surface.
At nearly full hydration two layers of water with higher than bulk
density are evident.
Few molecules
are trapped inside small pockets close to the
surface, which are a byproduct of the ``sample preparation process''.  
There is a strong tendency of water molecules
close to the surface to form hydrogen bonds
(HB) with the atoms of the substrate, 
in particular the hydrogens of water molecules with bO.
As a consequence the HB network of water
results to be strongly distorted close to the Vycor surface
\cite{spohr,spohr2}. This is 
compatible with the findings of recent 
experiments on water-in-Vycor \cite{mar1}. 

\section {SINGLE PARTICLE DYNAMICS OF CONFINED WATER}

Bulk SPC/E water, like
many glass forming liquid, when supercooled, or when the 
pressure is increased, develops a diversification of relaxation 
times scales, one fast and one slow, well described by MCT
\cite{gallo}. 
In the MCT \cite{goetze} scenario  a liquid approaches the glass transition 
point with a dynamic behaviour  mastered by the so called  ``cage effect''.  
The molecule or atom
is trapped by the transient cage formed by its nearest neighbour, $nn$.
After an initial ballistic regime it starts feeling the 
potential barrier of its $nn$. 
In this intermediate time region the particle is rattling in the cage
formed by this potential barrier.
When the cage relaxes the particle is free to move
and enters the normal diffusive regime. 
But as the liquid gets closer to $T_C$  the cage relaxation
time becomes longer and eventually approaches infinity in the idealized
MCT. Below $T_C$ cages are frozen and only hopping processes can restore
ergodicity allowing the particle to move (extended MCT) \cite{goetze2}.  
It is important here to stress that the cage effect is usually due in liquids 
to the increase of density, i.e. of $nn$, and in this respect water plays 
a peculiar role since there is no substantial increase in the 
number of $nn$ on  supercooling. 
It is the increased stiffness of the HB network 
at low temperature that allows for a ``cage'' to be formed 
also in bulk supercooled water.
\begin{figure}[t]
\centering\epsfig{file=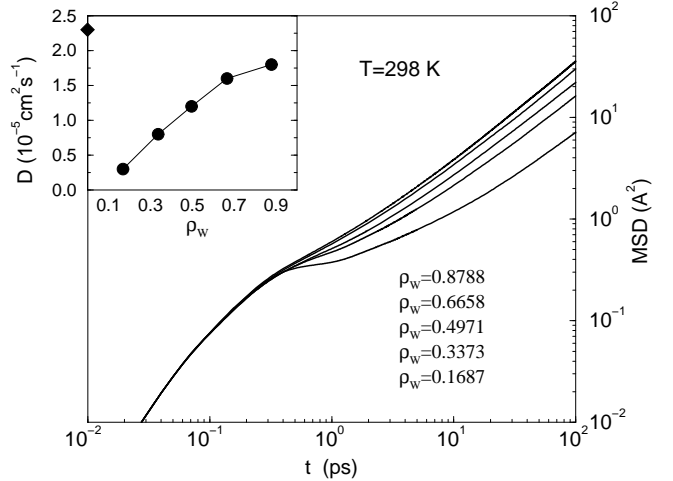,width=1\linewidth}
\caption{ MSD at ambient temperature of oxygens
in the non-confined z-direction
as a function of decreasing hydration from top to bottom.  
In the inset the diffusion
coefficients are shown. They correspond to the various 
hydration levels (circles) and to bulk SPC/E
water at ambient conditions (diamond).}
\protect\label{fig:3}
\end{figure} 

Signatures of a behaviour \`a la Mode Coupling can be 
found in all the correlators that have an overlap
with the density. In the following I will analyze for confined water
in particular 
the mean square displacement,
MSD, and the single particle density-density correlation function
in the Q,t space also called the intermediate scattering function, ISF, or
$F_S(Q,t)$. The signatures of this diversification
of relaxation times and therefore the shouldering of the relaxation
laws are more evident for the $Q$ values close to the peak of
the structure factor that for oxygen at ambient temperature is
$Q=2.25\AA^{-1}$. In particular the long time tail of the ISF is predicted
to have a stretched exponential behaviour when the system approaches the
glass transition.
\begin{figure}[t]
\centering\epsfig{file=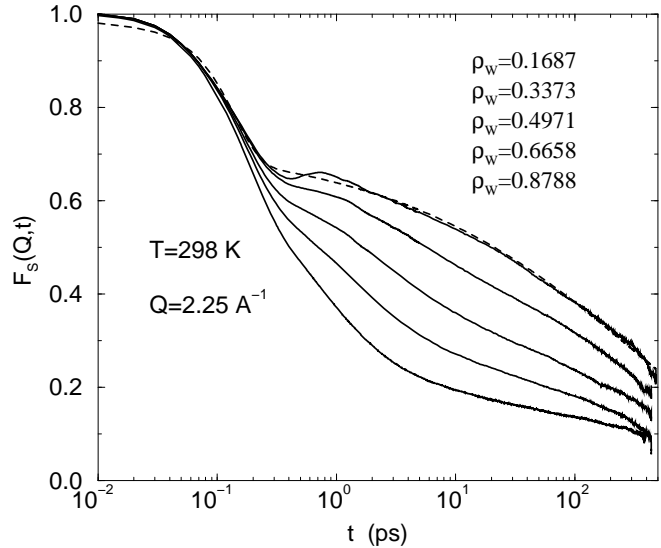,width=1\linewidth}
\caption{ISF of oxygens at
the peak of the structure factor
in the confined xy direction. 
The highly non-exponential long time tail 
could be fitted (dashed line) to eq.\ref{strexp}
only for the lower hydration.}
\protect\label{fig:4}
\end{figure} 

In Fig.3 the mean square displacement of oxygens
is displayed in the non-confined 
z-direction at ambient temperature as a function of  hydration level.
After the initial 
ballistic diffusion, before entering the
diffusive regime, a flattening of 
the curve at intermediate times is observed 
as the hydration level is decreased. In the inset the bulk diffusion 
coefficient and the confined ones are displayed. 
There is a substantial decrease
of average mobility as the hydration level in the pore is lowered. 
Correspondingly the oxygens ISF at the 
oxygen-oxygen peak of the structure factor, Fig.4,
displays a shouldering of
the relaxation laws upon decreasing hydration level. 
All the tails of the correlators of Fig.4
are highly non-exponential. None-the-less
these ISF could not be fitted
to the same formula used for bulk supercooled water
\cite{gallo}, 
\begin{equation}
F_S(Q,t)=
\left[ 1-A(Q) \right] e^{-\left( t/\tau_s \right)^2}+
A(Q)e^{- \left( t/\tau_l \right)^\beta}
\label{strexp}
\end{equation}
except for the lowest hydration.
In eq.\ref{strexp} $A(Q)=e^{-a^2Q^2/3}$ is the Lamb-M\"ossbauer factor 
(the analogous of the Debye Waller Factor for the single particle)
arising from the cage effect, 
and $\tau_s$ and $\tau_l$ are, respectively, the short and
the long relaxation times.
For the
fit of the ISF of $\rho_W=0.1687$ $g/cm^3$ shown in Fig.4
a cage radius $a\simeq0.44$ $\AA$ is obtained, which is similar
to the radius obtained for bulk supercooled water
where $a\simeq0.5$ \AA.
The lower radius obtained for confined water may be 
due to the slightly higher density of water close to the surface,
see Fig.2.
The short relaxation time $\tau_s\simeq 0.14$~ps is
again comparable to the bulk value which is $\tau_s\simeq 0.2$~ps.
$\tau_l\simeq356$~ps and $\beta=0.35$.
The $\beta$ associated with $\tau_l$ is very low compared to the 
typical values obtained for the supercooled bulk. Moreover
the lowest hydration correspond to less than the monolayer coverage so
that the dynamics of this system
is that of clusters of water molecules attached 
to the pore surface. 
For the lowest hydration it is also observed a bump around $0.7 ps$ that 
could be possibly related to the existence of the  Boson Peak feature 
in the $S(q,\omega)$ \cite{binder}. 
This point will be discussed at the end of the paragraph.
For the remaining correlators of Fig.4 no analytic function was found that
could fit the strongly non-exponential tails. 
Due to the strong hydrophilicity of the pore a diversification of 
dynamic behaviour is to be expected as we proceed from the pore surface to
the  center of the pore.
In Fig.5 the $F_S(Q,t)$ for the highest global density at ambient temperature 
is split into the contribution coming from the two layers of water 
molecules closer to the pore surface, outer shells, and into 
the contribution coming from all the remaining ones, inner shells.
The first two layers are defined according to the density profile,
Fig.2.
The inner shell contribution could be perfectly fit to eq.\ref{strexp}
as shown in the figure, while the outer shells one decays to zero over a 
much longer timescale so that water molecules there behave already as a 
glass.
From the fit it is extracted 
$\beta=0.74$, $\tau_l=0.75$ $ ps$ $\tau_s=0.21$ $ps$. The fit shows a 
remarkable agreement
and both the $\tau$ values are similar to those of bulk water,
while for bulk water at ambient temperature $\beta=1$. 
This kind of analysis has been 
extended for a different hydration and a different 
temperature. 
\begin{figure}[t]
\centering\epsfig{file=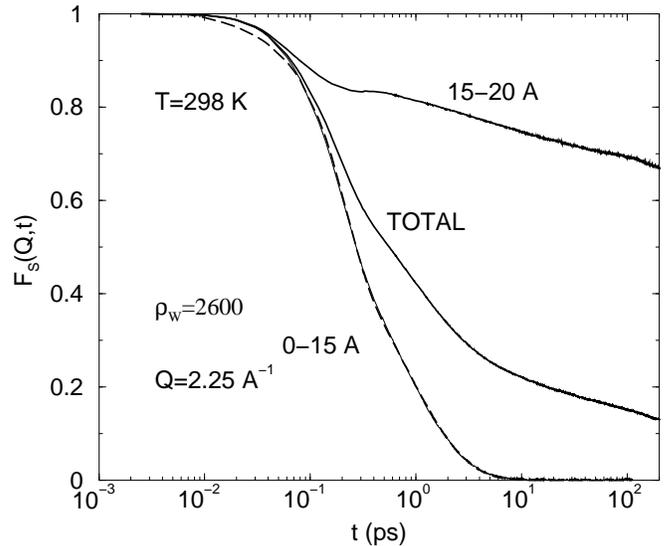,width=1\linewidth}
\caption{ISF for oxygens at almost full hydration
in the confined xy direction. Shell Analysis.
The top curve is the contribution to the total ISF coming from the molecules 
in the two shells closest to the substrate, see Fig.2. 
The bottom curve is the contribution
from the molecules in the remaining shells. The central one is the total ISF.
The dashed line is the fit to eq.\ref{strexp}.}
\protect\label{fig:5}
\end{figure} 
\begin{figure}[t]
\centering\epsfig{file=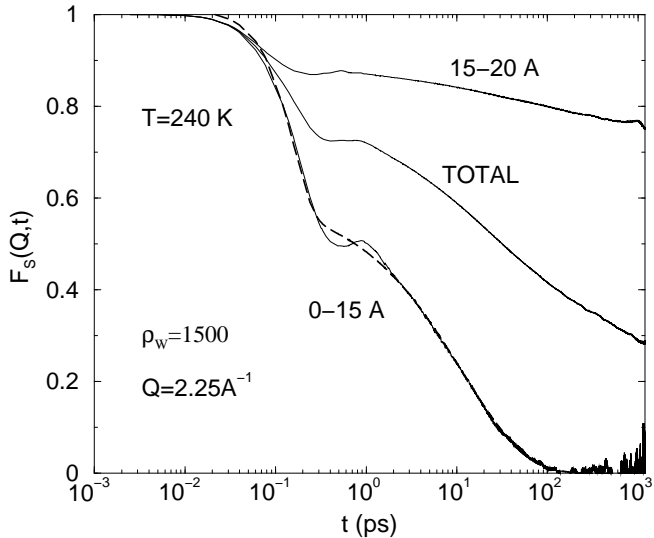,width=1\linewidth}
\caption{Shell analysis for the supercooled regime.
The dashed line is again the fit to eq. \ref{strexp}. 
Note that the bump related to the BP is now evident only for the
inner shell contribution.}
\protect\label{fig:6}
\end{figure} 
In Fig.6 the ISF shell analysis 
is shown at the peak of the structure factor for T=240 K and $\rho_W=0.4971$ $g/cm^3$.
Also in this case the fit to the stretched exponential of the tail
of the inner shells contribution is remarkable.
From the fit  
$\beta=0.62$, $\tau_l=11$ $ ps$ $\tau_s=0.16$ $ps$ are extracted. 
Again the  $\tau_l$ is similar to that of bulk water at the same
temperature while the $\beta$ here is much lower. 
In Fig.7 the $\tau_l$ and the $\beta$ values extracted from the
fits to eq.\ref{strexp} done as a function of $Q$ for T=240 K and 
$\rho_W=0.4971$ $g/cm^3$ are shown.
The $\beta$ value reaches a plateau value and the $\tau_l$ values show a
$Q^2$ dependence. Both these behaviours are typical of a glass former 
undergoing a kinetic glass transition and in particular the $Q^2$
behaviour has been observed
for example in glycerol close to the
glass transition \cite{glice}.

Let us now come back to Fig.4 where for the lowest hydration a bump
around 0.7 ps is clearly visible. Deeply supercooled bulk water
displayed the same peak around 0.35 ps \cite{gallo,fs}. 
This overshoot has been observed in 
simulations of supercooled strong glass formers like $SiO_2$ \cite{binder}.
It has been attributed in literature to
the existence of a Boson Peak (BP) in the frequency domain.
The BP is an excess of vibrational modes present in 
many glasses. When this glassy anomaly appears in the liquid
state it is  considered as precursor of the glass transition.
This peak in the ISF has been also alternatively
related to a disturbance propagating in a finite 
box. If there are periodic boundary conditions imposed (z direction
for the present system) the disturbance would reenter the box after
$t=L/v_s$ where $v_s$ is the sound velocity \cite{LW}.
In spite of the possible existence of this finite size effect 
it has been recently proven that the BP can 
be detected by MD \cite{binder}.
\begin{figure}[t]
\centering\epsfig{file=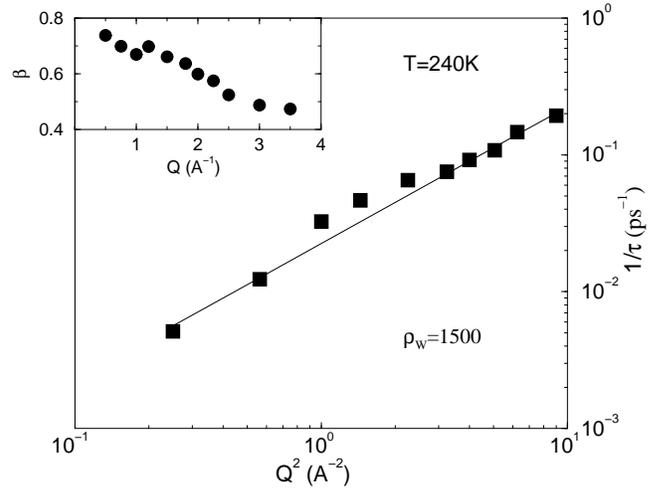,width=1\linewidth}
\caption{$\tau_l$ (main picture)
and $\beta$ (inset) values extracted from the 
fit of the long time tail of the ISF
to eq.\ref{strexp}. The continuous line has slope 2. 
The $Q^2$ behaviour has been found in glass formers close to the glass
transition.}
\protect\label{fig:7}
\end{figure} 
Note that in this system 
the same bump is present also in the supercooled confined 
simulation, see Fig.6, and it is more evident for the inner shells.
So that the presence of a disordered surface 
from one side seems to enhance the temperature at which the BP shows up
and also its intensity with respect to the bulk.
From the other side since  
the silica glass is rigid in our simulation the water molecules in
contact with the glass cannot vibrate much so that 
practically only the inner shells can
sustain vibrations. 
In Fig.8 same correlators of Fig.6 
in the xy and z direction are shown. It is important to stress here that
for this system the 
finite size effect in the xy directions is a feature
also of the real pore. 
\begin{figure}[t]
\centering\epsfig{file=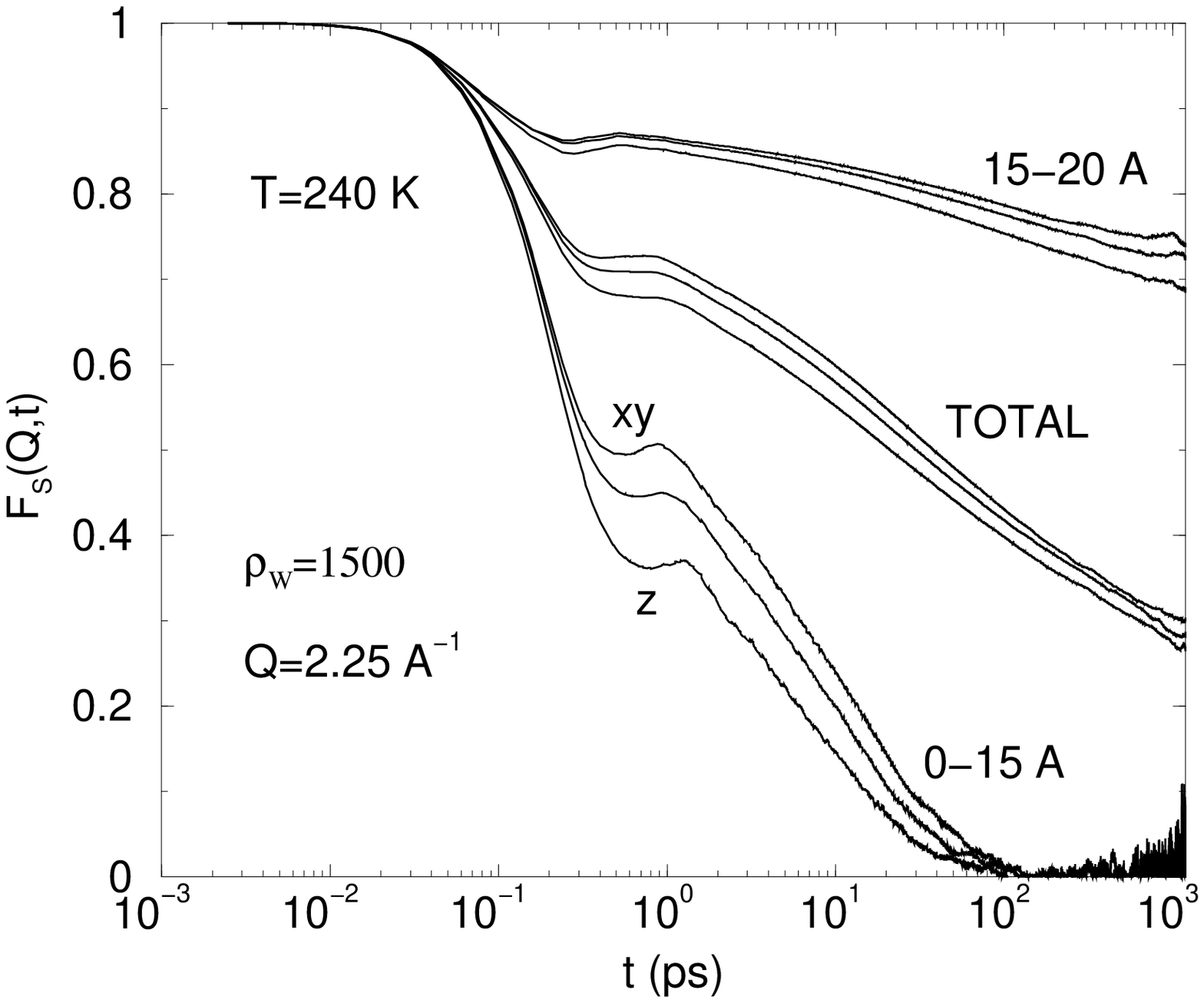,width=1\linewidth}
\caption{Shell analysis for the supercooled regime in
the confined (xy), non confined (z) and in an average 
random direction (central curve in each group of three). In particular:
ISF from the two shells closest to the substrate 
(top curves), inner shells 
(bottom curves) and total (central curves).}
\protect\label{fig:8}
\end{figure} 
This bump is therefore likely
to be observed for water-in-Vycor.
Both for the t corresponding
to the minima and the maxima of the bump appearing in the lowest 
curves of Fig.8 the ratio of the  
$t_z/t_{xy}$ is 1.44 against the 1.75 of the ratio between the
lengths of the pore $L_z=70\AA$ and $L_{xy}=40\AA$.
Moreover, although less evident, a BP can be observed also
for the outer shells in Fig.8
and the location is completely different with respect to
that of the inner shells. 
So that in our ISFs there is no direct quantitative connection 
with a finite size effect that can be detected from the shape 
and the location of these bumps. 
It can be inferred from these data that both the size and the geometry, 
and the presence of surface 
strongly influence the nature of this peak.
Size and geometry signatures can be seen in the 
change of location of the peak in the ISF in xy and z directions, see Fig.8. 
Signatures of a presence of a hydrophilic rigid surface 
can be related to the fact that  the bump is
more evident for the inner shells and is present already at room temperature.
The changes of location and shape with respect to the bulk
are therefore also probably 
influenced by all these causes. 
None-the-less, no quantitative statements can be made at this stage.
Therefore both a frequency domain analysis and a instantaneous normal mode 
analysis can help to shed light on this issue and they are currently 
in progress on this system.

\section {DISCUSSION AND CONCLUSIONS}

The results discussed in this paper show that 
for SPC/E water model an analogy between the supercooled bulk 
and the confined  as a function of hydration level of the pore
is possible only to the extent that in
confined water upon decreasing the hydration level a glassy behaviour 
appears already at ambient temperature while for bulk water
supercooling is required. None-the-less the manner the
confined liquid approaches the kinetic glass transition temperature appears 
completely different. 
In the pore at higher hydration levels, due to the hydrophilic surface, 
two quite distinct subsets of water molecules are detectable. 
Experiments in favour of two phases for confined hydrogen bonded 
liquids are present in literature \cite{melni}.
The subset 
that is in contact with the surface is at higher density
with respect to the bulk and is already a glass with low mobility
even at ambient temperature. The inner subset displays, like
a supercooled liquid, a two step relaxation behaviour \`a la MCT. 
It behaves none-the-less differently from the supercooled bulk. 
In fact it turns out that
confined and bulk relaxation times are comparable at the
same temperatures but the $\beta$ values 
of the confined water are much lower than those of the bulk
always at the same temperature. 
About the so called Kolhraush exponent 
$\beta$ can be said that in general it is
different for
different systems but
apparently no special significance can be attributed 
to its numerical value \cite{goetze}.
It is also important to mention here that SPC/E potential is considered one
of the best potential for water upon supercooling, but
these features are not produced by all potentials.
For example the ST2 shows
a ``jump diffusion'' behaviour upon supercooling 
\cite{Geiger}.
These two potential in particular are known to sandwich the behaviour 
of experimental water so that ST2 water
is less and SPC/E water is more structured than real water.
On the other hand experimental signatures of MCT
behaviour have been found both in the bulk \cite{bulkmct}
and in water-in-Vycor \cite{chen,chenla} and also 
experimental signatures of a
possible existence of a BP \cite{canni} in water-in-Vycor have been detected.
These results are encouraging and more simulations on this pore as
a function of temperature and hydration are in progress for a full
MCT test and a complete comparison with the bulk.

\section {Acknowledgements}

I wish to thank Mauro Rovere, Maria Antonietta Ricci and  Eckhard 
Spohr for their contributions to this work,
C. Austen Angell, Burkhard Geil, Alberto Robledo and Francesco
Sciortino for stimulating discussions on topics related to this paper.

\end{multicols}

\end{document}